\documentclass[proceedings]{JHEP3}

\usepackage[dvips]{graphicx}
\usepackage{amsmath}

\allowdisplaybreaks

\newcommand{\subref}[2] {\ref{#1}(#2)}

\newcommand{\eg} {\textit{e.g.}}

\newcommand{\LT}   {\left}
\newcommand{\RT}   {\right}

\newcommand{\CL}   {C.L.}
\newcommand{\dof}  {d.o.f.}
\newcommand{\eVq}  {\text{eV}^2}
\newcommand{\Sol}  {{\text{sol}}}
\newcommand{\Atm}  {{\text{atm}}}
\newcommand{\Dmq}  {\Delta m^2}
\newcommand{\Dms}  {\Delta m^2_\Sol}
\newcommand{\Dma}  {\Delta m^2_\Atm}
\newcommand{\Dcq}  {\Delta\chi^2}

\PrHEP{PrHEP AHEP2003/050}

\conference{International Workshop on Astroparticle and High Energy}

\title{Status of neutrino oscillations I:\\
  the three-neutrino scenario}

\author{\speaker{Michele Maltoni}\\
  IFIC, CSIC/Universitat de Val\`encia, Apt 22085, E--46071 Valencia,
  Spain
  \\
  YITP, SUNY at Stony Brook, Stony Brook, NY 11794-3840, USA
  \\
  E-mail: \email{maltoni@insti.physics.sunysb.edu}
  }

\abstract{
  We present a global analysis of neutrino oscillation data within the
  three-neutrino oscillation scheme, including in our fit all the
  current solar neutrino data, the reactor neutrino data from KamLAND
  and CHOOZ, the atmospheric neutrino data from Super-Kamiokande and
  MACRO, and the first data from the K2K long-baseline accelerator
  experiment. We determine the current best fit values and allowed
  ranges for the three-flavor oscillation parameters, discussing the
  relevance of each individual data set as well as the complementarity
  of different data sets. Furthermore, we analyze in detail the status
  of the small parameters $\theta_{13}$ and $\Dmq_{21} / \Dmq_{31}$,
  which fix the possible strength of CP violating effects in neutrino
  oscillations.
  }


\preprint{YITP-SB-04-67}

\begin{document}

\section{Introduction}

Recently, the Sudbury Neutrino Observatory (SNO)
experiment~\cite{Ahmed:2003kj} has released an improved measurement
with enhanced neutral current sensitivity due to neutron capture on
salt, which has been added to the heavy water in the SNO detector.
This adds precious information to the large amount of data on neutrino
oscillations published in the last few years. Thanks to this growing
body of data a rather clear picture of the neutrino sector is starting
to emerge. In particular, the results of the KamLAND reactor
experiment~\cite{Eguchi:2002dm} have played an important role in
confirming that the disappearance of solar electron
neutrinos~\cite{Cleveland:1998nv, Abdurashitov:2002nt, Hampel:1998xg,
Fukuda:2002pe, Ahmad:2002jz}, the long-standing solar neutrino
problem, is mainly due to oscillations and not to other types of
neutrino conversions. Moreover, KamLAND has pinned down that the
oscillation solution to the solar neutrino problem is the large mixing
angle MSW solution (LMA)~\cite{Maltoni:2002aw, Bahcall:2002ij,
Fogli:2002au, deHolanda:2002iv} characterized by the presence of
matter effects~\cite{Wolfenstein:1978ue, Mikheev:1985gs}.
On the other hand, experiments with atmospheric
neutrinos~\cite{Allison:1999ms, Shiozawa:2002, Surdo:2002rk} show
strong evidence in favor of $\nu_\mu \to \nu_\tau$ oscillations, in
agreement with the first data from the K2K accelerator
experiment~\cite{Ahn:2002up}. Together with the non-observation of
oscillations in reactor experiments at a baseline of about
1~km~\cite{Apollonio:1999ae, Boehm:2001ik}, and the strong rejection
against transitions involving sterile states~\cite{Maltoni:2002ni},
these positive evidences can be very naturally accounted for within a
three-neutrino framework. The large and nearly maximal mixing angles
indicated by the solar and atmospheric neutrino data samples,
respectively, come as a surprise for particle physicists, as it
contrasts with the small angles characterizing the quark sector.

In this talk we present a complete analysis of all the available
neutrino data in the context of the three-neutrino oscillation
scenario.
As a first step, in Sec.~\ref{sec:twonu} we discuss each individual
data set (solar, atmospheric, CHOOZ, KamLAND and K2K) in a simplified
two-neutrino approach. Then, in Sec.~\ref{sec:threenu} we discuss the
general three-neutrino case, focusing on the complementarity of
different data sets and and deriving the allowed ranges for all the
oscillation parameters. Furthermore, we discuss in detail the
constraints on the small parameters $\theta_{13}$ and the ratio
$\Dmq_{21} / \Dmq_{31}$, which are the crucial quantities governing
the possible strength of CP violating effects in neutrino
oscillations. In Sec.~\ref{sec:conclusions} we summarize our results.

\section{Two-neutrino analysis}
\label{sec:twonu}

\subsection{Solar neutrino data}
\label{sec:two.solar}

In order to determine the values of neutrino masses and mixing for the
oscillation solution of the solar neutrino problem, we have taken into
account the most recent results from all the solar neutrino
experiments. This includes the rate of the chlorine experiment at the
Homestake mine~\cite{Cleveland:1998nv} ($2.56 \pm 0.16 \pm 0.16$~SNU),
the latest results~\cite{taup2003} of the gallium experiments
SAGE~\cite{Abdurashitov:2002nt}
($69.1~^{+4.3}_{-4.2}~^{+3.8}_{-3.4}$~SNU) and
GALLEX/GNO~\cite{Hampel:1998xg} ($69.3 \pm 4.1 \pm 3.6$), the 1496-day
Super-Kamiokande data sample~\cite{Fukuda:2002pe} in the form of 44
bins (8 energy bins, 6 of which are further divided into 7 zenith
angle bins), the SNO-I neutral current, spectral and day/night
data~\cite{Ahmad:2002jz} divided into 34 bins (17 energy bins for each
day and night period), and the very recent SNO-II measurement of
charged current (CC), neutral current (NC) and elastic scattering (ES)
interaction rates with enhanced neutral current sensitivity due to
neutron capture on salt. Therefore, in our statistical analysis we use
$3 + 44 + 34 + 3 = 84$ observables, which we fit in terms of the two
parameters $\Dms$ and $\theta_\Sol$. The analysis methods used here
are similar to the ones described in Refs.~\cite{Maltoni:2002ni,
Gonzalez-Garcia:2000sq} and references therein, with the exception
that in the current work we use the so-called pull-approach for the
$\chi^2$ calculation. As described in Ref.~\cite{Fogli:2002pt}, each
systematic uncertainty is included by introducing a new parameter in
the fit and adding a penalty function to the $\chi^2$. However, the
method described in Ref.~\cite{Fogli:2002pt} is extended in two
respects. First, it is generalized to the case of correlated
statistical errors~\cite{Balantekin:2003jm}, as necessary to treat the
SNO-salt data. Second, in the calculation of the total $\chi^2$ we use
the exact relation between the theoretical predictions and the pulls
associated to the solar neutrino fluxes, rather than keeping only the
terms up to first order. This is particularly relevant for the case of
the solar $^8$B flux, which is constrained by the new SNO data with an
accuracy better than the prediction of the Standard Solar Model
(SSM)~\cite{Bahcall:1998wm}. In our approach it is possible to take
into account on the same footing both the SSM boron flux prediction
and the SNO NC measurement, without pre-selecting one particular
value. In this way \textit{the fit itself} can choose the best
compromise between the SNO NC data and the SSM value.

In Fig.~\subref{fig:solar-atmos}{a} we show the allowed regions and
the $\Dcq$ functions from the analysis of all solar neutrino
experiments, including the new SNO-salt data. One finds that
especially the upper part of the LMA region and large mixing angles
are now strongly constrained~\cite{Maltoni:2003da}. This follows
mainly from the rather small measured value of the CC/NC ratio of
$0.306 \pm 0.026 \pm 0.024$~\cite{Ahmed:2003kj}, since this observable
increases when moving to larger values of $\Dms$ and/or
$\sin^2\theta_\Sol$ (see, \eg, Ref.~\cite{deHolanda:2002iv}). In is
worth noting that the LOW solution is now excluded at more than
$3\sigma$, and the quasi-vacuum region at more than $5\sigma$, even
without the inclusion of KamLAND. The best fit values for the
oscillation parameters are
\begin{equation}\label{eq:bfp-solar}
    \sin^2\theta_\Sol = 0.29 \,,\qquad
    \Dms = 6.0 \times 10^{-5}~\eVq \qquad\text{(solar data)}.
\end{equation}
The new SNO-salt data strongly enhances the rejection against maximal
solar mixing: from the $\Dcq$ projected onto the $\sin^2\theta_\Sol$
axis, shown in Fig.~\subref{fig:solar-atmos}{a}, we find that
$\theta_\Sol = 45^\circ$ is excluded at more than $5\sigma$, ruling
out all bi-maximal models of neutrino masses.

\FIGURE[t]{
  \includegraphics[scale=0.47]{fig.solar-atmos.eps}
  \caption{\label{fig:solar-atmos}%
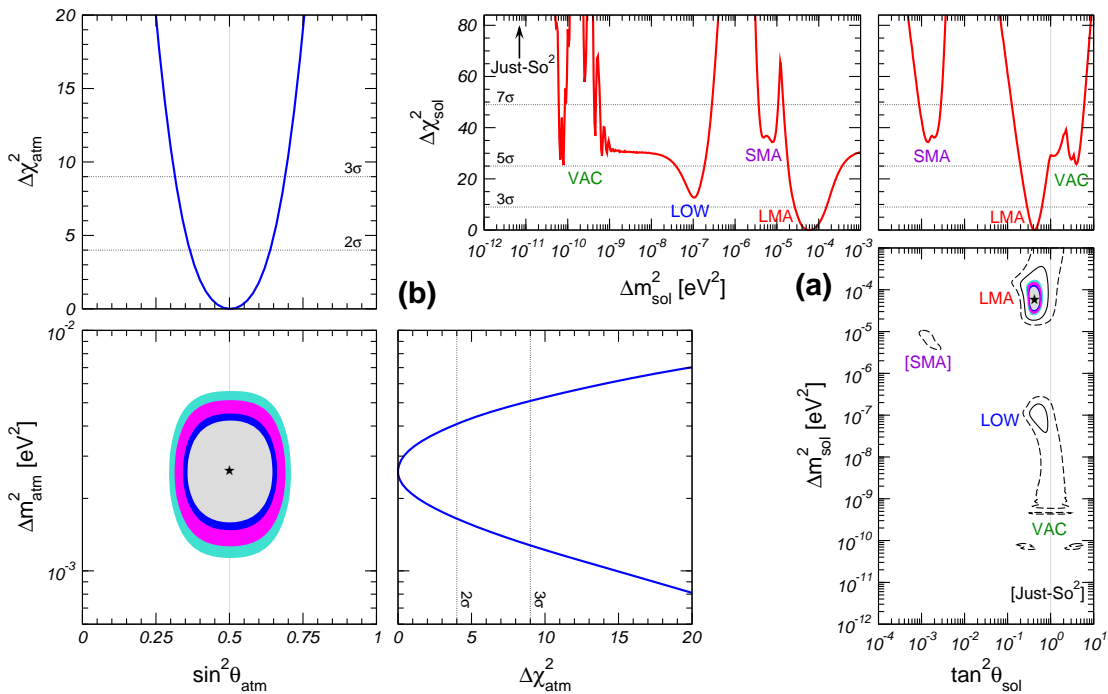
    Allowed regions at 90\%, 95\%, 99\%, and $3\sigma$ \CL\ (2 \dof)
    from the analysis of solar neutrino data (a) and of atmospheric
    neutrino data (b), in the two-neutrino oscillation approximation.
    In the solar plane (a) we also show the $5\sigma$ (solid line) and
    $7\sigma$ (dashed line) allowed regions. Also displayed is $\Dcq$
    as a function of the relevant oscillation parameters.}
  }

\subsection{Atmospheric neutrino data}
\label{sec:two.atmos}

For the atmospheric data analysis we use all the charged-current data
from the Super-Kamiokande~\cite{Shiozawa:2002} and
MACRO~\cite{Surdo:2002rk} experiments. The Super-Kamiokande data
include the $e$-like and $\mu$-like data samples of sub- and multi-GeV
contained events (10 bins in zenith angle), as well as the stopping (5
angular bins) and through-going (10 angular bins) up-going muon data
events. As previously, we do not use the information on $\nu_\tau$
appearance, multi-ring $\mu$ and neutral-current events since an
efficient Monte-Carlo simulation of these data sample would require a
more detailed knowledge of the Super Kamiokande experiment, and in
particular of the way the neutral-current signal is extracted from the
data. Such an information is presently not available to us. From MACRO
we use the through-going muon sample divided in 10 angular
bins~\cite{Surdo:2002rk}. We did not include in our fit the results of
other atmospheric neutrino experiments since at the moment the
statistics is completely dominated by Super-Kamiokande.

Our statistical analysis of the atmospheric data described in
Ref.~\cite{Maltoni:2002ni, Gonzalez-Garcia:2000sq}. In particular, we
now take advantage of the full ten-bin zenith-angle distribution for
the contained events, rather than the five-bin distribution employed
in our older publications. Therefore, we have now $65$ observables,
which we fit in terms of the two relevant parameters $\Dma$ and
$\theta_\Atm$. Concerning the theoretical Monte-Carlo, we improved the
method presented in Ref.~\cite{Gonzalez-Garcia:2000sq} by properly
taking into account the scattering angle between the incoming neutrino
and the scattered lepton directions. This was already the case for
Sub-GeV contained events, however previously we made the simplifying
assumption of full neutrino-lepton collinearity in the calculation of
the expected event numbers for the Multi-GeV contained and
up-going-$\mu$ data samples. While this approximation is still
justified for the stopping and through-going muon samples, in the
Multi-GeV sample the theoretically predicted value for down-coming
$\nu_\mu$ is systematically higher if full collinearity is assumed.
The reason for this is that the strong suppression observed in these
bins cannot be completely ascribed to the oscillation of the
down-coming neutrinos (which is small due to small travel distance).
Because of the non-negligible neutrino-lepton scattering angle at
these Multi-GeV energies there is a sizable contribution from up-going
neutrinos (with a higher conversion probability due to the longer
travel distance) to the down-coming leptons.
However, this problem is less visible when the angular information of
Multi-GeV events is included in a five angular bins presentation of
the data, as previously assumed.

We note that recently the Super-Kamiokande collaboration has
presented a preliminary reanalysis of their atmospheric
data~\cite{sk:aachen}. The update includes changes in the detector
simulation, data analysis and input atmospheric neutrino fluxes. These
changes lead to a slight downward shift of the mass-splitting to the
best fit value of $\Dma = 2\times 10^{-3}~\eVq$. Currently it is not
possible to recover enough information from Ref.~\cite{sk:aachen} to
incorporate the corresponding changes in our codes. We note, however,
that the quoted value for $\Dma$ is statistically compatible with our
result. For $\Dma = 2\times 10^{-3}~\eVq$ and maximal mixing we obtain
a $\Dcq = 1.3$.

In Fig.~\subref{fig:solar-atmos}{b} we summarize our results for the
analysis of atmospheric neutrino experiments. First of all, we note
that the dependence of the atmospheric $\Dcq$ over the oscillation
parameters $\Dma$ and $\theta_\Atm$ exhibits a beautiful quadratic
behavior, reflecting the fact that the oscillation solution to the
atmospheric neutrino problem is robust and unique. The best fit values
and the $3\sigma$ ranges (1 \dof) for the oscillation parameters are:
\begin{align}
    \sin^2\theta_\Atm &= 0.5 \,, 
    & 0.3 \le \sin^2\theta_\Atm & \le 0.7 \,,
    \\
    \Dma &= 2.5 \times 10^{-3}~\eVq \,,
    & 1.2 \times 10^{-3}~\eVq \le \Dma &\le 4.8 \times 10^{-3}~\eVq\,.
\end{align}
Note that in the two-neutrino approximation the neutrino conversion
occur completely in the channel $\nu_\mu \to \nu_\tau$, and in this
case the contour regions are symmetric under the transformation
$\theta_\Atm \to \pi/4 - \theta_\Atm$ due to the cancellation of
matter effects.

\subsection{CHOOZ}
\label{sec:two.chooz}

\FIGURE[t]{
  \includegraphics[scale=0.47]{fig.react-accel.eps}
  \caption{\label{fig:react-accel}%
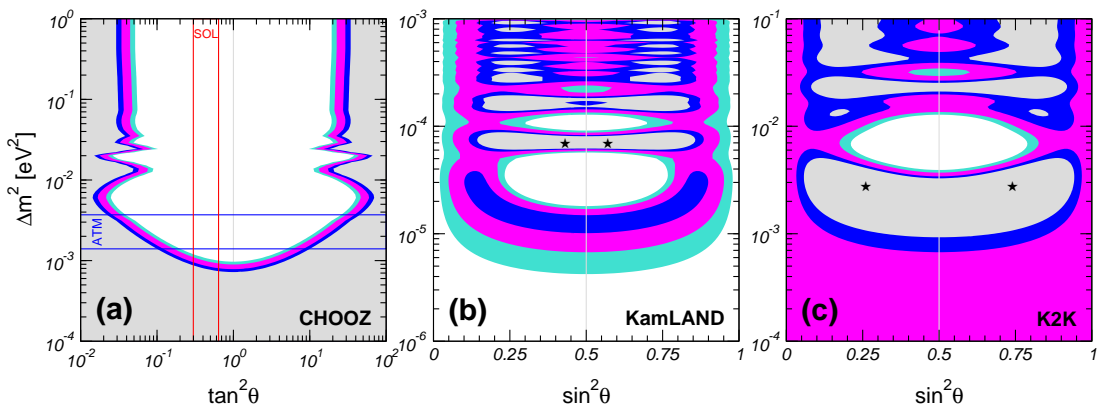
    Allowed regions at 90\%, 95\%, 99\%, and $3\sigma$ \CL\ (2 \dof)
    from the analysis of CHOOZ (a), KamLAND (b), and K2K data (c).}
  }

The CHOOZ experiment~\cite{Apollonio:1999ae} searches for
disappearance of $\bar{\nu}_e$ produced in a power station with two
pressurized-water nuclear reactors with a total thermal power of
$8.5$~GW (thermal). At the detector, located at $L\simeq 1$~Km from
the reactors, the $\bar{\nu}_e$ reaction signature is the delayed
coincidence between the prompt ${\rm e^+}$ signal and the signal due
to the neutron capture in the Gd-loaded scintillator. Since no
evidence was found for a deficit of measured vs.\ expected neutrino
interactions, the result of this experiment is a bound on the
oscillation parameters describing the mixing of $\nu_e$ with all the
other neutrino species. In the two-neutrino limit the only relevant
parameters to describe the oscillations are a mass-squared splitting
$\Dmq$ and a mixing angle $\theta$, and in
Fig.~\subref{fig:react-accel}{a} we show the exclusion plot in this
plane. From this figure we see that at the $3\sigma$ level the region
$\Dmq \gtrsim 10^{-3}~\eVq$ and $0.05 \lesssim \tan^2\theta \lesssim
20$ is excluded.

The result of this experiment, once combined with solar and
atmospheric data, has important implications for the determination of
the oscillations parameters. This can be qualitatively understood even
without performing a detailed three-neutrino analysis. For what
concerns solar data, since solar neutrino conversion occur in the
channel $\nu_e \to \nu_a$, which is the same channel measured by CHOOZ
once the CPT symmetry is assumed, the constraints imposed by CHOOZ on
the solar parameters can be immediately derived from
Fig.~\subref{fig:react-accel}{a} by identifying $\Dmq$ and $\theta$
with $\Dms$ and $\theta_\Sol$, respectively. The two vertical red
lines in Fig.~\subref{fig:react-accel}{a} delimit the $3\sigma$
allowed region for $\theta_\Sol$, as found in
Sec.~\ref{sec:two.solar}, and it is straightforward to see that values
of $\Dms \gtrsim 10^{-3}~\eVq$ are incompatible with the CHOOZ result.
Therefore, CHOOZ implies an upper bound on $\Dms$. This result was
particularly relevant before the release of the SNO data on
neutral-current interactions, since at that time the CHOOZ bound was
the only measurement providing a solid upper bound on $\Dms$.

Concerning atmospheric data, the implications of the CHOOZ result
require a more complete formalism. In the next section we will see
that, in the three-neutrino scenario, the mixing of the electron
neutrino with the $\nu_\mu-\nu_\tau$ admixture responsible for
atmospheric conversion is described by a mixing angle which we will
denote by $\theta_{13}$. Therefore, the two-dimensional exclusion plot
shown if Fig.~\subref{fig:react-accel}{a} can also be interpreted as a
constraint in the plane $(\Dma, \, \theta_{13})$. The two horizontal
blue lines in this figure delimit the $3\sigma$ allowed region in
$\Dma$, as found in Sec.~\ref{sec:two.atmos}, and we can see that the
CHOOZ measurement implies that $\theta_{13}$ must be small. This means
that the electron neutrino is essentially decoupled from atmospheric
neutrino oscillations, thus justifying \emph{a posteriori} the use of
the two-neutrino approximation for the analysis of atmospheric data.

\subsection{KamLAND}

The KamLAND experiment is a reactor neutrino experiment with its
detector located at the Kamiokande site. Most of the $\bar{\nu}_e$
flux incident at KamLAND comes from plants at distances of $80-350$ km
from the detector, making the average baseline of about 180
kilometers, long enough to provide a sensitive probe of the LMA
solution of the solar neutrino problem.
In KamLAND the target for the $\bar{\nu}_e$ flux consists of a
spherical transparent balloon filled with 1000 tons of non-doped
liquid scintillator, and the anti-neutrinos are detected via the
inverse neutron $\beta$-decay $\bar{\nu}_e+p \to e^{+}+n$.

The KamLAND collaboration has for the first time measured the
disappearance of neutrinos traveling to a detector from a power
reactor.
They observe a strong evidence for the disappearance of electron
anti-neutrinos during their flight, giving the first terrestrial
confirmation of the solar neutrino anomaly and also establishing the
oscillation hypothesis with man-produced neutrinos.
Moreover, the parameters that describe this disappearance of the
electron neutrino in terms of oscillations are perfectly consistent
with latest determinations of solar neutrino parameters in a
CPT-conserving scenario.

The details of the statistical analysis of the KamLAND data can be
found in Ref.~\cite{Schwetz:2003se}. Instead of the usual $\chi^2$-fit
analysis based on energy binned data, we use an event-by-event
likelihood approach which gives stronger constraints. The results of
our analysis are summarized in Fig.~\subref{fig:react-accel}{b}, where
we show the allowed regions of the oscillation parameters. It is in
good agreement with the analysis performed by the KamLAND group, shown
in Fig.~6 of Ref.~\cite{Eguchi:2002dm}. This gives us confidence on
our simulation of the KamLAND data and therefore encourages us to use
it in a full analysis combining also with the other data samples.

\subsection{K2K}

The last data set which we include in our fits is the KEK to Kamioka
long-baseline neutrino oscillation experiment (K2K)~\cite{Ahn:2002up}.
The neutrino beam used by this experiment is composed by muon
neutrinos at 98\%, and has a mean energy of 1.3~GeV. Since the
neutrino flight distance is approximately 250~km, K2K can provide a
sensitive proves of atmospheric oscillations. In our analysis we use
the 29 single-ring muon events, grouped into 6 energy bins, and we
perform a spectral analysis similar to the one described in
Ref.~\cite{Fogli:2003th}. Our results are summarized in
Fig.~\subref{fig:react-accel}{c}.

\section{Global three-neutrino analysis}
\label{sec:threenu}

\subsection{Notation}

In general, the determination of the oscillation probabilities
requires the solution of the Schr\"odinger evolution equation of the
neutrino system in the Sun- and/or Earth-matter
background~\cite{Fogli:2002au, Gonzalez-Garcia:2000sq,
  Gonzalez-Garcia:2003qf}:
\begin{equation}
    i \frac{d\vec{\nu}}{dt} = {\bf H} \, \vec{\nu}, \qquad
    {\bf H} = {\bf U} \cdot {\bf H}_0^d \cdot {\bf U}^\dagger + {\bf V},
\end{equation}
where ${\bf U}$ is the unitary matrix connecting the flavor basis and the
mass basis in vacuum, ${\bf H}_0^d$ is the vacuum Hamiltonian and {\bf V}
describes charged-current forward interactions in matter. For a CP-conserving
three-flavor scenario, we have:
\begin{equation} \label{eq:evol}
    \begin{aligned}
	{\bf H}_0^d &= \frac{1}{2 E_\nu} \, {\bf diag}
	\LT( - \Dmq_{21},\, 0,\, \Dmq_{32} \RT);
	\\
	{\bf U} &= {\bf U}_{23}(\theta_{23}) 
	\cdot {\bf U}_{13}(\theta_{13})	\cdot {\bf U}_{12}(\theta_{12})\,;
	\\
	{\bf V} &= \pm \sqrt{2} G_F \, {\bf diag} \LT( N_e,\, 0,\, 0 \RT);
	\\
	\vec\nu &= \LT( \nu_e,\, \nu_\mu,\, \nu_\tau \RT).
    \end{aligned}
\end{equation}
In the general case, the three-neutrino transition probabilities
depend on five parameters, namely the two mass-squared differences
$\Dmq_{21}$, $\Dmq_{31}$ and the three mixing angles $\theta_{12}$,
$\theta_{13}$, $\theta_{23}$. However, from the two-neutrino analysis
presented in the previous section we see that, in order to accommodate
both solar and atmospheric neutrino data into a unified framework, an
hierarchy between the two mass-squared differences is required:
\begin{equation}
    \LT( \Dms \equiv \Dmq_{21} \RT) \ll
    \LT( \Dma \equiv \Dmq_{31} \approx \Dmq_{32} \RT).
\end{equation}
Both for the solar and the atmospheric neutrino data analysis we can
take advantage of this hierarchy to reduce the number of
parameters~\cite{Gonzalez-Garcia:2002mu}. This implies first of all
that the effect of a possible Dirac CP-violating phase in the lepton
mixing matrix can be neglected. Also, for what concerns the solar case
we can set $\Dmq_{31} \approx \infty$ and also disregard the
atmospheric angle $\theta_{23}$, since it never appears in the
relevant transition probabilities. Conversely, for the atmospheric
case we can set $\Dmq_{21} \approx 0$ and in this limit the solar
angle $\theta_{12}$ cancels out from the equations. So in both the
cases we are left with only three parameters, among which only the
reactor angle $\theta_{13}$ is common to both the problems. We will
further comment on the goodness of the hierarchy approximation at the
end of the next section.

\subsection{Discussion}

\FIGURE[t]{
  \includegraphics[scale=0.47]{fig.solkam-atmk2k.eps}
  \caption{\label{fig:solkam-atmk2k}%
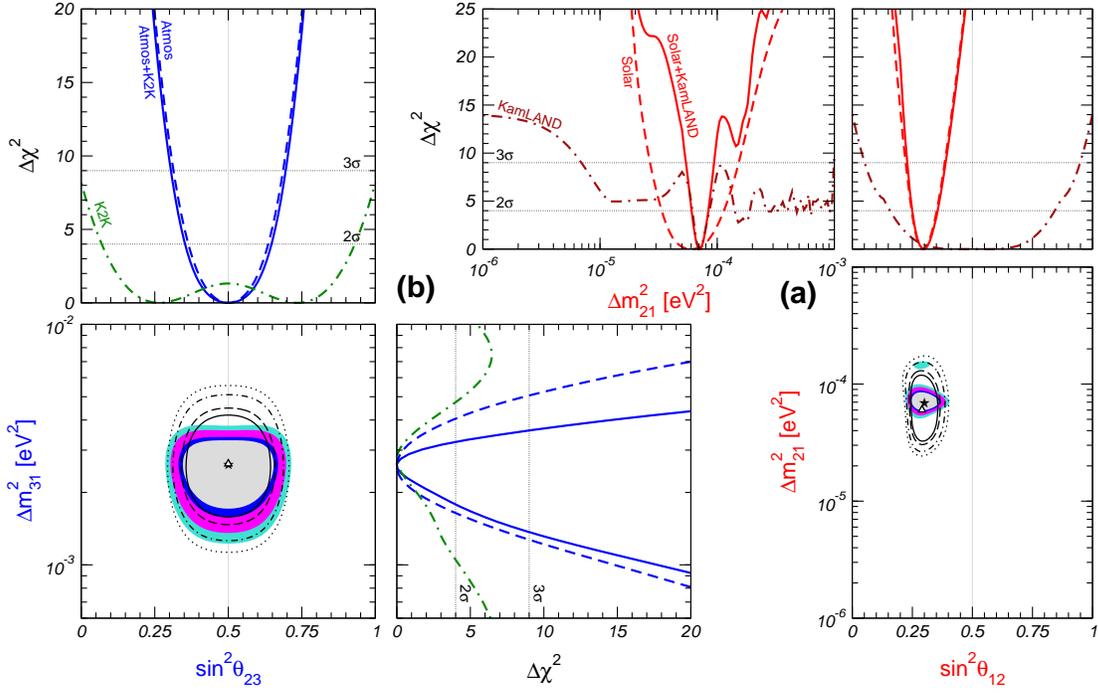
    Allowed regions at 90\%, 95\%, 99\%, and $3\sigma$ \CL\ (2 \dof)
    from the analysis of solar (hollow regions) and solar+KamLAND
    (colored regions) neutrino data (a), and of atmospheric (hollow
    regions) and atmospheric+K2K (colored regions) neutrino data (b),
    in terms of three-neutrino oscillation parameters. In both
    analyses we set $\theta_{13} = 0$. Also displayed is $\Dcq$ as a
    function of the relevant oscillation parameters.}
  }

Having discussed each independent data set separately, we now turn our
attention to the bound on neutrino oscillation parameters coming from
combinations of different data sets. The results of our analysis are
summarized in Figs.~\ref{fig:solkam-atmk2k} and \ref{fig:global}. In
Figs.~\subref{fig:solkam-atmk2k}{a} and \subref{fig:solkam-atmk2k}{b}
we show the projections on the planes $(\Dmq_{21},\, \theta_{12})$ and
$(\Dmq_{31},\, \theta_{23})$ of the combination of solar+KamLAND data
and atmospheric+K2K data, respectively, for $\theta_{13} = 0$. In
Fig.~\ref{fig:global} we show the allowed regions and the $\Dcq$
function from the analysis of all neutrino data.

\FIGURE[t]{
  \includegraphics[scale=0.5]{fig.global.eps}
  \caption{\label{fig:global}%
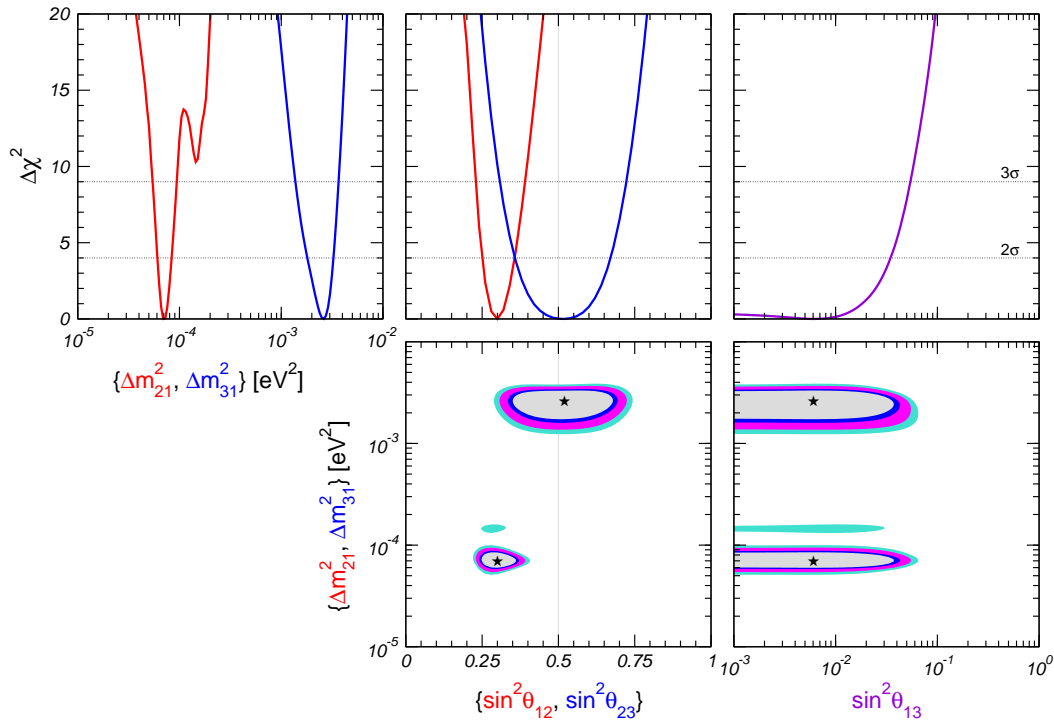
    Projections of the allowed regions from the global oscillation
    data at 90\%, 95\%, 99\%, and $3\sigma$ \CL\ for 2 \dof\ for
    various parameter combinations. Also shown is $\Dcq$ as a function
    of the oscillation parameters $\theta_{12}$, $\theta_{23}$,
    $\theta_{13}$, $\Dmq_{21}$ and $\Dmq_{31}$, minimized with respect
    to all undisplayed parameters.}
  }

From Figs.~\subref{fig:solkam-atmk2k}{a} and
\subref{fig:solkam-atmk2k}{b}, we see that the inclusion of KamLAND
and K2K data drastically improves the determination of the two
mass-squared differences $\Dmq_{21}$ and $\Dmq_{31}$. In particular,
K2K is responsible for the strong improvement of the upper bound on
$\Dmq_{31}$, leaving the lower bound essentially unaffected. On the
other hand, the recent KamLAND result single out LMA as the only
viable oscillation solution to the solar neutrino problem. As noted in
Ref.~\cite{Maltoni:2002aw}, the original LMA region is split by
KamLAND into two separate islands, called LMA-I (characterized by
$\Dmq_{21} \approx 7 \times 10^{-5}~\eVq$) and LMA-II (with $\Dmq_{21}
\approx 1.5 \times 10^{-4}~\eVq$). The inclusion of the recent
SNO-salt data practically rules out the LMA-II solution, which is now
disfavored with a $\Dcq = 10.3$ with respect to
LMA-I~\cite{Maltoni:2003da}. The best-fit point and the $3\sigma$
ranges for the global analysis (see Fig.~\ref{fig:global}) are:
\begin{align}
    \Dmq_{21} & = 6.9 \times 10^{-5}~\eVq \,,
    & 5.4 \times 10^{-5}~\eVq \leq \Dmq_{21} & \leq 9.5 \times 10^{-5}~\eVq \,,
    \\
    \Dmq_{31} & = 2.6 \times 10^{-3}~\eVq \,,
    & 1.4 \times 10^{-3}~\eVq \leq \Dmq_{31} & \leq 3.7 \times 10^{-3}~\eVq \,.
\end{align}

The determination of the oscillation angles $\theta_{12}$ and
$\theta_{23}$ is dominated by solar and atmospheric data,
respectively, while reactor and accelerator experiments only play a
marginal role in this respect. As discussed in
Sec.~\ref{sec:two.solar}, the recent SNO-salt data drastically
improves the upper bound on the solar angle $\theta_{12}$, and maximal
mixing $\theta_{12} = 45^\circ$ is now ruled out at more than
$5\sigma$. Conversely, atmospheric data clearly prefer $\theta_{23} =
45^\circ$. From the global analysis, we have that best-fit point and
the $3\sigma$ ranges are:
\begin{align}
    \sin^2\theta_{12} &= 0.3 \,,
    & 0.23 \leq \sin^2\theta_{12} & \leq 0.39 \,,
    \\
    \sin^2\theta_{23} & = 0.52 \,,
    & 0.36 \leq \sin^2\theta_{23} & \leq 0.67 \,.
\end{align}
Note that the small deviation from maximal mixing of $\theta_{23}$ is
due to the fact that the data prefer a small but non-vanishing value
of $\theta_{13}$, which breaks the invariance under the transformation
$\theta_{23} \to \pi/2 - \theta_{23}$. However, currently this effect
has no statistical significance at all.

\FIGURE[t]{
  \includegraphics[scale=0.47]{fig.subleading.eps}
  \caption{\label{fig:subleading}%
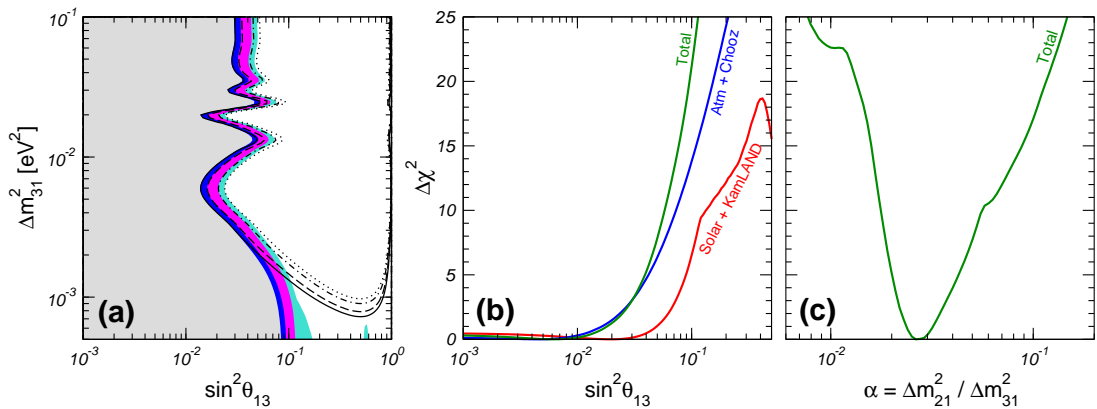
    (a) Allowed regions in the $(\sin^2\theta_{13},\, \Dmq_{31})$ plane at
    90\%, 95\%, 99\%, and $3\sigma$ from CHOOZ data alone (lines) and
    CHOOZ+solar+KamLAND data (colored regions). (b) $\Dcq$ profiles
    projected onto the $\sin^2\theta_{13}$ axis, for solar+KamLAND,
    atmospheric+CHOOZ, and for global data. (c) $\Dcq$ from global
    oscillation data as a function of $\alpha \equiv \Dmq_{21} /
    \Dmq_{31}$.}
  }

The last angle in the three-neutrino mixing matrix, $\theta_{13}$, is
presently still unknown. At the moment only an upper bound exists,
which is mainly dominated by the CHOOZ reactor
experiment~\cite{Apollonio:1999ae}.
In Fig.~\subref{fig:subleading}{b} we show the $\Delta\chi^2$ as a function
of $\sin^2\theta_{13}$ for different data sample choices. One can see
how the bound on $\sin^2\theta_{13}$ as implied by the CHOOZ
experiment in combination with the atmospheric neutrino data still
provides the main restriction on $\sin^2\theta_{13}$. We find the
following bounds at 90\% \CL\ ($3\sigma$) for 1 \dof:
\begin{equation}\label{eq:th13}
    \sin^2\theta_{13} \le \left\{ \begin{array}{l@{\qquad}l}
      0.070~(0.12) & \text{(solar+KamLAND)} \\
      0.028~(0.066) & \text{(CHOOZ+atmospheric)} \\
      0.029~(0.054) & \text{(global data)}
  \end{array} \right.
\end{equation}

However, we note that the solar data contributes in an important way
to the constraint on $\sin^2\theta_{13}$ for lower values of
$\Dmq_{31}$. In particular, the downward shift of $\Dmq_{31}$ reported
in Ref.~\cite{sk:aachen} implies a significant loosening of the CHOOZ
bound on $\sin^2\theta_{13}$, since this bound gets quickly weak when
$\Dmq_{31}$ decreases (see, \eg, Ref.~\cite{Fogli:2003am}). Such
loosening in sensitivity is prevented to some extent by solar neutrino
data. In Fig.~\subref{fig:subleading}{a} we show the allowed regions
in the $(\sin^2\theta_{13},\, \Dmq_{31})$ plane from an analysis
including solar and reactor neutrino data (CHOOZ and KamLAND). One
finds that, although for larger $\Dmq_{31}$ values the bound on
$\sin^2\theta_{13}$ is dominated by the CHOOZ + atmospheric data, for
low $\Dmq_{31} \simeq 10^{-3}~\eVq$ the solar + KamLAND bound is
comparable to that coming from CHOOZ + atmospheric. For example,
fixing $\Dmq_{31} = 2\times 10^{-3}~\eVq$ we obtain at 90\%
\CL\ ($3\sigma$) for 1 \dof\ the bound
\begin{equation}
    \sin^2\theta_{13} \le 0.035~(0.066) 
    \qquad (\Dmq_{31} = 2\times 10^{-3}~\eVq) \,,
\end{equation}
which is slightly worse than the bound from global data shown in
Eq.~\eqref{eq:th13}.

For the exploration of genuine three-flavor effects such as
CP-violation the mass hierarchy parameter $\alpha \equiv \Dmq_{21} /
\Dmq_{31}$ is of crucial importance since, in a three-neutrino scheme,
CP violation disappears in the limit where two neutrinos become
degenerate. Therefore we show in Fig.~\subref{fig:subleading}{c} the
$\Delta\chi^2$ from the global data as a function of this parameter.
We obtain the following best fit values and $3\sigma$ intervals:
\begin{equation} \begin{split}
    \alpha = 0.026\,, \qquad 0.018 \le \alpha \le 0.053.
\end{split} \end{equation}
Note that the strong upper bound found on $\alpha$ justifies \emph{a
posteriori} the use of the hierarchy approximation in the calculation
of solar and atmospheric $\chi^2$
functions~\cite{Gonzalez-Garcia:2002mu}.

\section{Conclusions}
\label{sec:conclusions}

In this talk we have presented a global analysis of neutrino
oscillation data in the three-neutrino scheme, including in our fit
all the current solar neutrino data, the reactor neutrino data from
KamLAND and CHOOZ, the atmospheric neutrino data from
Super-Kamiokande and MACRO, and the first data from the K2K
long-baseline accelerator experiment. We have discussed the
implications of each individual data set, as well as the
complementarity of different data sets, on the determination of the
neutrino oscillation parameters, and we have determined the current
best fit values and allowed ranges for the three-flavor oscillation
parameters $\theta_{12}$, $\theta_{23}$, $\theta_{13}$, $\Dmq_{21}$,
$\Dmq_{31}$.
Furthermore, we have analyzed in detail the limits on the small mixing
angle $\sin^2\theta_{13}$ and on the hierarchy parameter $\alpha =
\Dmq_{21} / \Dmq_{31}$. These small parameters are relevant for
genuine three flavor effects, and restrict the magnitude of leptonic
CP violation that one may potentially probe at future experiments like
super beams or neutrino factories. In particular, we have seen how the
improvement on the $\theta_{13}$ limit that follows from the new
SNO-salt data solar neutrino experiments can not yet match the
sensitivity reached at reactor experiments. However, the solar data do
play an important role in stabilizing the constraint on $\theta_{13}$
with respect to variation of $\Dmq_{31}$.  For small enough
$\Dmq_{31}$ values the solar data probe $\theta_{13}$ at a level
comparable to that of the current reactor experiments.

\acknowledgments

Talk based on the work performed in collaboration with M.C.\
Gonzalez-Garcia, T.\ Schwetz, M.A.\ T\'ortola and J.W.F.\ Valle.  This
work was supported by Spanish grant BFM2002-00345, by the European
Commission RTN network HPRN-CT-2000-00148, by the European Science
Foundation network grant N.~86 and by the National Science Foundation
grant PHY0098527.

\end{document}